\documentclass[reprint, showpacs, superscriptaddress, english, twocolumn, notitlepage, longbibliography, nofootinbib]{revtex4-1}
\usepackage[T1]{fontenc}
\newcommand\thefontsize{\expandafter\string\the\font}

\usepackage[utf8]{inputenc}
\pdfoutput=1
\usepackage{times}
\usepackage{color}
\usepackage{array}
\usepackage{verbatim}
\usepackage{multirow}
\usepackage{amsmath}
\usepackage{amssymb}
\usepackage{dsfont}
\usepackage{graphicx}
\usepackage{esint}
\usepackage[unicode=true,
 bookmarks=true,bookmarksnumbered=true,bookmarksopen=false,
 breaklinks=true,pdfborder={0 0 0},backref=false,colorlinks=true]
 {hyperref}
\usepackage{enumitem}
\usepackage{titlesec}
\usepackage{cleveref}
\usepackage[normalem]{ulem}

\hypersetup{
    linkcolor=red,          
    citecolor=blue,        
    filecolor=magenta,      
    urlcolor=magenta           
}

\crefname{appendix}{Sec.}{Secs.}
\crefname{equation}{Eq.}{Eqs.}
\crefname{figure}{Fig.}{Figs.}
\crefname{table}{Table}{Tables}
\crefname{section}{Sec.}{Secs.}

\renewcommand{\paragraph}[1]{\vspace{0.2cm}{\bf \textit{#1}}}
\def\ie{i.e.,\ }
\def\eg{e.g.,\ }
\def\etc{etc.\ }

\definecolor{Gray}{gray}{0.85}
\newcolumntype{a}{>{\columncolor{Gray}}c}

\usepackage{simpler-wick}
\allowdisplaybreaks[1] %

\def\pare#1{\left( #1 \right)}
\def\brak#1{\left[#1\right]}
\def\brace#1{\left\{#1\right\}}
\def\bra#1{\langle #1 |}
\def\ket#1{| #1 \rangle}
\def\Bra#1{\left\langle #1 \right|}
\def\Ket#1{\left| #1 \right\rangle}
\def\inn#1{\langle #1 \rangle}
\def\Inn#1{\left\langle #1 \right\rangle}
\def\abs#1{\left| #1 \right|}
\def\Im{\mathrm{Im}}
\def\Re{\mathrm{Re}}

\def\mF{\mathcal{F}}
\def\ii{\mathrm{i}}
\def\pp{\mathbf{p}}
\def\Tr{\mathrm{Tr}}
\def\kk{\mathbf{k}}
\def\GG{\mathbf{G}}
\def\mA{\mathcal{A}}
\def\mD{\mathcal{D}}
\def\RR{\mathbf{R}}
\def\tt{\mathbf{t}}

\begin{document}
\title{Orbital Magnetization of Interacting Electrons}

\author{Xi Chen}
\affiliation{International Center for Quantum Materials, School of Physics, Peking University, Beijing 100871, China}

\author{Zhi-Da Song}
\email{songzd@pku.edu.cn}
\affiliation{International Center for Quantum Materials, School of Physics, Peking University, Beijing 100871, China}
\affiliation{Hefei National Laboratory, Hefei 230088, China}
\affiliation{Collaborative Innovation Center of Quantum Matter, Beijing 100871, China}

\date{\today}

\begin{abstract}
We derive an exact expression for the orbital magnetization of electrons with short-range interactions (such as density-density interactions) in terms of exact zero-frequency response functions of the zero-field system.
The result applies to weakly and strongly correlated electrons at zero and finite temperature, provided that the local grand potential density only depends on local thermodynamic parameters.
We benchmark the formula for non-interacting and weakly-coupled electrons.
To zeroth and first orders in the interaction strength, it agrees with the modern theory of orbital magnetization and its recent generalization to self-consistent Hartree-Fock bands.
Our work provides an exact framework of interacting orbital magnetization beyond mean-field treatments, and paves the way for quantitative studies of strongly correlated electrons in external magnetic fields.
\end{abstract}

\maketitle

\textbf{\textit{Introduction.}}
Orbital magnetization (OM)~\cite{Thonhauser2005OMPRL,Ceresoli2006OMPRB,Xiao2010BerryPhaseRMP,doi:10.1142/S0217979211058912} arises from the orbital motion of itinerant electrons in systems with broken time-reversal symmetry.
While typically weaker than spin magnetization in conventional materials, it has recently attracted significant attention following the experimental realization of spontaneous ferromagnetic states in moir\'e superlattices ~\cite{Sharpe2019EmergentFM,Lu2019OrbitalMagnets,Serlin2020QAH,Chen2020TunableChern,Tschirhart2021ImagingOrbitalFM,LiuDai2021MoireOrbitalMagnetism}.
In these systems, the orbital moment can be electrically tuned to rival or exceed the spin contribution ~\cite{Tschirhart2021ImagingOrbitalFM,LiuDai2021MoireOrbitalMagnetism,Liu2025TMDHFOM,Song2021MoireMagnetismDomains,He2021tMBGOrbitalMagnetism,Yu2022CorrelatedHofstadterMATBG,xie2025unconventionalorbitalmagnetismgraphenebased}, driving novel phenomena and necessitating a rigorous theoretical framework for OM in the presence of strong electron--electron interactions.

The quantum mechanical treatment of OM under periodic boundary conditions is subtle because the magnetization operator $\hat{\boldsymbol{M}} = (-e/2\hbar)\,\hat{\boldsymbol{r}}\times \hat{\boldsymbol{v}}$ involves the unbounded position operator $\hat{\boldsymbol{r}}$~\cite{Thonhauser2005OMPRL,Shi2007QuantumOM}.
For non-interacting electrons, this difficulty is resolved by the modern theory, which expresses OM as a bulk Brillouin-zone quantity relevant to the quantum geometry of Bloch bands~\cite{Thonhauser2005OMPRL,Ceresoli2006OMPRB,Xiao2005BerryPhaseDOS,Xiao2010BerryPhaseRMP,Shi2007QuantumOM}.
This result is supported by complementary approaches, including semiclassical wave-packet dynamics~\cite{SundaramNiu1999Wavepacket,Xiao2005BerryPhaseDOS,Xiao2010BerryPhaseRMP}, real-space Wannier formulations~\cite{Thonhauser2005OMPRL,Ceresoli2006OMPRB,Lopez2012WannierOM,doi:10.1142/S0217979211058912}, and gauge invariant Green's-function methods~\cite{ChenLee2011Unified,PhysRevB.91.085120,Zhu2012DisorderedOM}.
One particular approach ~\cite{Shi2007QuantumOM} is to compute the linear grand potential density response to a weak magnetic field slowly varying in space, and then take the long-wavelength limit to recover the case of a uniform magnetic field.
This approach exploits the key observation that, although the operator $\hat{\boldsymbol{M}}$ is non-local, the linear response of the grand potential density depends only on the local magnetic field, provided that the spatial modulation of the field is sufficiently slow.

Despite the success of the modern theory, its generalization to interacting electrons remains a long-standing challenge.
The formalism has been successfully generalized to static mean-field descriptions, including current and spin density functional theory~\cite{Shi2007QuantumOM,VignaleRasolt1987CSDFT,VignaleRasolt1988CSDFT}, and most recently, self-consistent Hartree-Fock theory~\cite{Kang2025HFOM,Liu2025TMDHFOM,ZhuHuang2025Projectors}.
For strongly correlated electrons, an interacting Green's-function expression was derived and implemented~\cite{Nourafkan2014DMFTOM,PhysRevB.99.075144,ZhouPandeyFeng2021VI3} within Dynamical Mean-Field Theory (DMFT)~\cite{Georges1996DMFT}, where the linear response of the gauge-invariant (local) self-energy to the magnetic field is neglected.
While this is a reasonable approximation for DMFT, it is not straightforwardly generalizable to systems with significant non-local correlations.
Consequently, it would be highly valuable to derive a general exact formula for the OM of electrons with strong density-density interactions, expressed entirely in terms of zero-field quantities.

In this work, we derive an exact formula of OM in terms of exact response functions.
By including infinitely slowly spatial modulations of both the magnetic field and the local energy scale, the calculation of OM can be reformulated as a standard perturbative problem.
Then using the Lehmann spectral representation \cite{AltlandSimons2010CMFT}, the OM is written in closed form of zero-frequency response functions of the zero-field unperturbed system.
The formula requires certain correlation functions to decay faster than $r^{-2}$ at long distances.
This condition quantitatively captures the assumed locality of the grand potential density response and justifies applying our formalism to a broad class of interacting systems, including (but not limited to) insulators and Fermi liquids in two and three spatial dimensions.
The modern theory is readily recovered in the non-interacting limit, and a first order weak-coupling expansion agrees with recent results on self-consistent Hartree-Fock calculations ~\cite{Kang2025HFOM,Liu2025TMDHFOM,ZhuHuang2025Projectors}.
Our approach provides an exact treatment of OM for correlated electrons beyond mean field theories, and paves the way for the development of controlled approximation schemes for its accurate calculation for strongly correlated materials.

\textbf{\textit{Model.}}
We consider a tight-binding model on a periodic lattice with density-density interactions,
\begin{equation}
\small
\begin{aligned}
    \hat{K}_0 = \hat{H}-\mu \hat{N} =& \sum_{\boldsymbol{R_1}\boldsymbol{R_2}\alpha\beta}h_{\alpha\beta}(\boldsymbol{R_1}-\boldsymbol{R_2})c_{\boldsymbol{R_1},\alpha}^{\dagger}c_{\boldsymbol{R_2},\beta} \\&+ V_{\alpha\beta}(\boldsymbol{R_1} - \boldsymbol{R_2})c_{\boldsymbol{R_1},\alpha}^{\dagger}c_{\boldsymbol{R_2},\beta}^{\dagger}c_{\boldsymbol{R_2},\beta}c_{\boldsymbol{R_1},\alpha}
\end{aligned}
\label{eq:Hamiltonian}
\end{equation}
where $\boldsymbol{R}$ labels unit cells, and $\alpha,\beta$ label orbitals located at $\boldsymbol{R}+\boldsymbol{r}_{\alpha,\beta}$. $h_{\alpha\beta}(\boldsymbol{\Delta R})$ and $V_{\alpha\beta}(\boldsymbol{\Delta R})$ denote the bilinear hoppings and the density-density interactions, respectively, both of which are assumed to be exponentially short-ranged. Note that we have absorbed the chemical potential $\mu$ into the definition of $h_{\alpha\beta}$.
A necessary requirement of our formalism is that the Peierls substitution induced by the magnetic field leaves the interaction term invariant. 
Without loss of generality, we assume density-density interactions for simplicity.

The OM at temperature $T$ is defined as $M_z =-\frac{\partial \Omega}{\partial B_z}|_{T, \mu}$ \cite{Shi2007QuantumOM}, where $\Omega = E-TS-\mu N = -\beta^{-1} \ln \Xi$ is the grand thermodynamic potential, $\beta = 1/(k_BT)$, and $\Xi = \sum_{i} e^{-\beta K_i}$ is the grand partition function of the system in an external magnetic field $B_z$. The sum runs over all exact many-body eigenstates $i$ with eigenvalues $K_i$ in the presence of $B_z$.

\textbf{\textit{Construction of an auxiliary system.}}
In Ref.~\cite{Shi2007QuantumOM}, an infinitely slowly oscillating magnetic field is applied to a uniform non-interacting system.
Assuming the {\it locality} of the grand potential density response, the OM is related to the appropriate Fourier component of grand potential density.
For interacting electrons, however, the grand potential density is in general very difficult to compute.
We therefore employ the locality condition to incorporate fluctuations of both the magnetic field and the local energy scale.
For clarity, we illustrate the formalism by constructing an auxiliary system.

Without loss of generality, we introduce the smallest momentum along $x$ direction compatible with periodic boundary conditions of linear size $L$, $\boldsymbol{q} = q\hat{\boldsymbol{x}}$, where
$q\sim 1/L$ is infinitesimal in the thermodynamic limit.
System A is defined by placing the many-body Hamiltonian \cref{eq:Hamiltonian} at temperature $T$, and it serves as the reference system of the perturbations.
We then construct an auxiliary spatially non-uniform Hamiltonian
\begin{equation}
\small
\begin{aligned}
    &\hat{K}_B = \sum_{\boldsymbol{R_1}\boldsymbol{R_2}\alpha\beta}[1+2\eta \cos(\boldsymbol{q}\cdot \frac{\boldsymbol{R_1+R_2+r_{\alpha}+r_{\beta}}}{2})]\\&\{h_{\alpha\beta}(\boldsymbol{R_1}-\boldsymbol{R_2})c_{\boldsymbol{R_1},\alpha}^{\dagger}c_{\boldsymbol{R_2},\beta}\exp(-\ii \frac{e}{\hbar} \int_{\boldsymbol{R_2+r_{\beta}}}^{\boldsymbol{R_1+r_{\alpha}}}\boldsymbol{A(r)\cdot dr}) \\
    &+ V_{\alpha\beta}(\boldsymbol{R_1} - \boldsymbol{R_2})c_{\boldsymbol{R_1},\alpha}^{\dagger}c_{\boldsymbol{R_2},\beta}^{\dagger}c_{\boldsymbol{R_2},\beta}c_{\boldsymbol{R_1},\alpha}\}
\end{aligned}
\label{eq:KB}
\end{equation}
where $\eta \ll 1$ is an artificial infinitesimal quantity, and we choose the vector potential $\boldsymbol{A}(\boldsymbol{r}) = A_0 \sin (\boldsymbol{q\cdot r}) \boldsymbol{\hat{y}}$, generating a magnetic field $\boldsymbol{B}(\boldsymbol{r}) = B_0 \cos(\boldsymbol{q\cdot r})\boldsymbol{\hat{z}}$, where $A_0$ is also infinitesimal such that $B_0 = qA_0$ is infinitesimal.
System B is defined by placing $\hat{K}_B$ at temperature $T$.
Physically, if $h_{\alpha\beta}$ and $V_{\alpha\beta}$ are both exponentially short-ranged, system B can be understood as a spatially nonuniform version of the base Hamiltonian $(1+2\eta \cos(\boldsymbol{q\cdot r}))\hat{K}_A$ subjected to the nonuniform magnetic field $\boldsymbol{B}(\boldsymbol{r})$.

We now evaluate the grand potential density of system B, $\Omega_B (\boldsymbol{r})$.
Since the spatial modulation of the inhomogeneous parameters occurs only at a macroscopic length scale $1/q\sim L$, it is reasonable to assume that $\Omega_B(\boldsymbol{r})$ only depends on the local base Hamiltonian, magnetic field, and temperature.
At each given position $\boldsymbol{r}$, we find that system B is equivalent to subjecting the base Hamiltonian \cref{eq:Hamiltonian} to magnetic field $\boldsymbol{B}(\boldsymbol{r})$ at temperature $[1+2\eta \cos(\boldsymbol{q\cdot r})]^{-1}T$, and then scaling all involved energy scales by a common factor $1+2\eta \cos(\boldsymbol{q\cdot r})$.
Since the grand potential also has dimensions of energy, its density should scale accordingly, so we have
{\small
\begin{equation}
\begin{aligned}
\Omega_B(\boldsymbol{r}) =& \mathcal{V}^{-1}(1+2\eta \cos(\boldsymbol{q\cdot r}))
 \Big\{\Omega_A -2\eta T\frac{\partial \Omega}{\partial T}\cos(\boldsymbol{q\cdot r}) \\
- &  B_0M_z(T) \cos(\boldsymbol{q\cdot r}) + 2\eta B_0 \cos^2(\boldsymbol{q\cdot r})T\frac{\partial M_z}{\partial T}|_\mu \Big\}
\end{aligned}
\end{equation}}
where $\mathcal{V}$ is the volume (area) of the system. 
We have neglected terms of order $O(\eta^2), O(B^2)$ in the derivation and made use of $\partial^2\Omega/\partial B\partial T = -\partial M_z/\partial T| _\mu$.
Integrating over $\boldsymbol{r}$ for $\Omega_B = \int d\boldsymbol{r}\, \Omega_B(\boldsymbol{r})$, we obtain
\begin{equation}
    \delta \Omega_B = - \eta B_0 M_z +\eta B_0 T\frac{\partial M_z}{\partial T}|_{\mu}
\end{equation}
where $\delta \Omega_B = \Omega_B - \Omega_A$. It then follows that
\begin{equation}
    -\frac{\partial^2 \delta\Omega_B}{\partial \eta\partial B_0} = \tilde{M}_z(T) = M_z(T) - T \frac{\partial M_z}{\partial T}|_\mu
\label{eq:connection_of_free_energy_to_OM}
\end{equation}
$\tilde{M}_z(T)$ also appears in Ref.~\cite{Shi2007QuantumOM}, where it is called the ``auxiliary magnetization'' and is derived from a different context.
At zero temperature it equals the proper OM $M_z$, and at finite temperature the two are related by
\begin{equation}
    M_z(T) = k_BT\int_{0}^{1/(k_BT)} d\beta\tilde{M}_z(\frac{1}{k_B \beta})
\end{equation}

On the other hand, \cref{eq:KB} can be written as $\hat{K_B} = \hat{K}_0 + \Delta \hat{K}_B$, where $\Delta \hat{K}_B$ is controlled by infinitesimal quantities $\eta$ and $A_0$. So $\delta \Omega_B$ can be directly expressed as
\begin{equation}
\delta \Omega_B = -\frac{1}{\beta} \Xi^{-1}\delta \Xi = \Xi^{-1} \sum_{i} \delta K_i e^{-\beta K_i}
\label{eq:Omega_B_perturb}
\end{equation}
where $\beta = 1/(k_BT)$ and $\delta K_i$ is the change of the $i^{\rm th}$ eigenvalue of $\hat{K}$ due to $\Delta \hat{K}_B$, which can be evaluated by standard perturbation theory. We now evaluate $\delta K_i$ in real space.

\textbf{\textit{Perturbative calculation in real space.}}
Expanding \cref{eq:KB} to first order in $\eta$, $A_0$, and $\eta A_0$, we obtain
\begin{widetext}
\begin{equation}
\small
\begin{aligned}
    \hat{\Delta K_B} &=2\eta \sum_{\boldsymbol{R_1R_2}, \alpha\beta}  h_{\alpha\beta}(\boldsymbol{R_1} - \boldsymbol{R_2}) \cos(\boldsymbol{q}\cdot \boldsymbol{R}^{C}_{12, \alpha\beta}) c_{\boldsymbol{R_1}, \alpha}^\dagger c_{\boldsymbol{R_2}, \beta}+ 2\eta \sum_{\boldsymbol{R_1R_2}, \alpha\beta}  V_{\alpha\beta}(\boldsymbol{R_1} - \boldsymbol{R_2}) \cos(\boldsymbol{q}\cdot \boldsymbol{R}^{C}_{12, \alpha\beta}) c_{\boldsymbol{R_1}, \alpha}^\dagger c_{\boldsymbol{R_2}, \beta}^\dagger c_{\boldsymbol{R_2}, \beta} c_{\boldsymbol{R_1}, \alpha}\\
    &-\frac{\ii e}{\hbar} A_0\sum_{\boldsymbol{R_1R_2}, \alpha\beta} h_{\alpha\beta}(\boldsymbol{R_1} - \boldsymbol{R_2}) (y_{\boldsymbol{R}_1, \alpha} - y_{\boldsymbol{R}_2, \beta}) \sin(\boldsymbol{q}\cdot \boldsymbol{R}^{C}_{12, \alpha\beta}) c_{\boldsymbol{R_1}, \alpha}^\dagger c_{\boldsymbol{R_2}, \beta}\\&-\frac{\ii e}{\hbar} \eta A_0\sum_{\boldsymbol{R_1R_2}, \alpha\beta} h_{\alpha\beta}(\boldsymbol{R_1} - \boldsymbol{R_2}) (y_{\boldsymbol{R}_1, \alpha} - y_{\boldsymbol{R}_2, \beta}) \sin(2\boldsymbol{q}\cdot \boldsymbol{R}^{C}_{12, \alpha\beta})
    c_{\boldsymbol{R_1}, \alpha}^\dagger c_{\boldsymbol{R_2}, \beta}\\
\end{aligned}
\label{eq:Delta_KB}
\end{equation}
\end{widetext}
where we have denoted $\boldsymbol{R}^C_{12, \alpha\beta} = \frac 1 2(\boldsymbol{R_1} + \boldsymbol{R_2} + \boldsymbol{r_\alpha} + \boldsymbol{r_\beta})$.

In general, $\hat{\Delta K_B}$ can shift $K_i$ through both degenerate and nondegenerate perturbation theory.
However, \cref{eq:Delta_KB} contains only terms carrying momentum $\pm \boldsymbol{q}$ and $\pm 2\boldsymbol{q}$, so its expectation value in any eigenstate of $\hat{K_0}$ vanishes.
Consequently, for any degenerate sector $\mathcal{S}$ of $\hat{K_0}$, we have $\Tr \hat{\Delta K_B}|_\mathcal{S} = 0$, and hence $\sum_{i\in \mathcal{S}} \delta K_i = 0$.
Since all $\delta K_i$ in sector $\mathcal{S}$ are multiplied by the same factor $e^{-\beta K_i}$ in \cref{eq:Omega_B_perturb}, the net contribution of degenerate perturbations to $\delta \Omega_B$ is zero.
The first-order energy shifts in nondegenerate perturbation theory also vanish by momentum conservation, while third- and higher-order perturbations unavoidably include $O(\eta^2)$ or $O(A_0^2)$ terms, which are irrelevant in \cref{eq:connection_of_free_energy_to_OM}. Therefore, it suffices to consider only second-order nondegenerate perturbations for our purpose.

For simplicity, we will denote the average of the retarded and advanced response functions as the principal-part contribution of the response functions,
$C^{\mathcal{P}}_{\hat{A}\hat{B}}(\omega ) = \frac 1 2(C^{r}_{\hat{A}\hat{B}}(\omega ) + C^{a}_{\hat{A}\hat{B}}(\omega ))$,
then the Lehmann spectral representation at zero frequency reads~\cite{AltlandSimons2010CMFT}
\begin{equation}
\small
    C^{\mathcal{P}}_{\hat{A}\hat{B}}(\omega=0) =  \Xi ^{-1} \sum_{ij} e^{-\beta K_i} (A_{ij}B_{ji}+ B_{ij}A_{ji}) \mathcal{P}(\frac{1}{K_i-K_j})
\label{eq:Lehmann_rep_zerofreq}
\end{equation}
provided that $\hat{A}$ and $\hat{B}$ are bosonic operators, where $\mathcal{P}$ denotes the Cauchy principal value.
~\cref{eq:Lehmann_rep_zerofreq} corresponds to the sum of all connected Feynman diagrams with external legs corresponding to $\hat{A}$ and $\hat{B}$ at zero (bosonic Matsubara) frequency.
Substituting the standard second-order perturbation formula for $\delta K_i$ to \cref{eq:Omega_B_perturb}, we obtain
\begin{equation}
    \delta \Omega_B = \frac 1 2 C^{\mathcal{P}}_{\hat{\Delta K_B}, \hat{\Delta K_B}} (\omega = 0)
\label{eq:Omega_diff_as_response}
\end{equation}
Substituting \cref{eq:Delta_KB} into \cref{eq:Omega_diff_as_response} and neglecting $O(\eta^2)$ and $O(A_0^2)$ terms, we find
\begin{widetext}
\begin{equation}
\small
\begin{aligned}
    \delta \Omega_B = -\frac{2\ii e}{\hbar} \eta A_0 \sum_{\boldsymbol{R_1...R_4}}\sum_{\alpha_1...\alpha_4} &h_{\alpha_1\alpha_2}(\boldsymbol{R_1-R_2}) (y_{\boldsymbol{R}_1, \alpha_1} - y_{\boldsymbol{R}_2, \alpha_2})
    \sin(\boldsymbol{q}\cdot \boldsymbol{R}^C_{12, \alpha_1\alpha_2}) \cos(\boldsymbol{q}\cdot \boldsymbol{R}^C_{34, \alpha_3\alpha_4})\\
    &[h_{\alpha_3\alpha_4}(\boldsymbol{R_3-R_4}) \mathcal{I}_{\alpha_1\alpha_2\alpha_3\alpha_4}(\boldsymbol{R_1, R_2, R_3, R_4})
    + V_{\alpha_3\alpha_4}(\boldsymbol{R_3-R_4}) \mathcal{J}_{\alpha_1\alpha_2\alpha_3\alpha_4} (\boldsymbol{R_1, R_2, R_3, R_4})]
\end{aligned}
\label{E_cossin}
\end{equation}
where we have denoted 
\begin{equation}
\small
    \mathcal{I}_{\alpha_1\alpha_2\alpha_3\alpha_4}(\boldsymbol{R_1, R_2, R_3, R_4}) = C^{\mathcal{P}}_{c_{\boldsymbol{R_1}, \alpha_1}^\dagger c_{\boldsymbol{R_2}, \alpha_2}, c_{\boldsymbol{R_3}, \alpha_3}^\dagger c_{\boldsymbol{R_4}, \alpha_4}}(\omega = 0)
\end{equation}
\begin{equation}
\small
    \mathcal{J}_{\alpha_1\alpha_2\alpha_3\alpha_4}(\boldsymbol{R_1, R_2, R_3, R_4}) = C^{\mathcal{P}}_{c_{\boldsymbol{R_1}, \alpha_1}^\dagger c_{\boldsymbol{R_2}, \alpha_2}, c_{\boldsymbol{R_3}, \alpha_3}^\dagger c_{\boldsymbol{R_4}, \alpha_4}^\dagger c_{\boldsymbol{R_4}, \alpha_4} c_{\boldsymbol{R_3}, \alpha_3}}(\omega = 0)
\end{equation}
\end{widetext}
Using $\sin (x)\cos(y) = \frac 1 2[\sin(x+y) + \sin(x-y)]$ together with the translational invariance of $\mathcal{I}$ and $\mathcal{J}$, the factor $\sin(\boldsymbol{q}\cdot \boldsymbol{R}^C_{12, \alpha_1\alpha_2}) \cos(\boldsymbol{q}\cdot \boldsymbol{R}^C_{34, \alpha_3\alpha_4})$ in \cref{E_cossin} can be replaced by $\frac 1 2 \sin(\boldsymbol{q}\cdot (\boldsymbol{R}^C_{12, \alpha_1\alpha_2}-\boldsymbol{R}^C_{34, \alpha_3\alpha_4}))$.
If we further assume the correlations to be sufficiently short-ranged (as discussed later), we may expand
$\sin(\boldsymbol{q}\cdot (\boldsymbol{R}^C_{12, \alpha_1\alpha_2}-\boldsymbol{R}^C_{34, \alpha_3\alpha_4})) \approx \boldsymbol{q}\cdot (\boldsymbol{R}^C_{12, \alpha_1\alpha_2}-\boldsymbol{R}^C_{34, \alpha_3\alpha_4})$
in the long-wavelength limit.
Substituting \cref{E_cossin} into \cref{eq:connection_of_free_energy_to_OM}, we obtain the auxiliary OM
\begin{widetext}
\begin{equation}
\small
\begin{aligned}
    \tilde{M}_z = \frac{\ii e}{2\hbar} \sum_{\boldsymbol{R_1...R_4}}\sum_{\alpha_1...\alpha_4} &h_{\alpha_1\alpha_2}(\boldsymbol{R_1-R_2}) (y_{\boldsymbol{R}_1, \alpha_1} - y_{\boldsymbol{R}_2, \alpha_2})
    (x_{\boldsymbol{R_1}, \alpha_1} + x_{\boldsymbol{R_2}, \alpha_2} - x_{\boldsymbol{R_3}, \alpha_3} - x_{\boldsymbol{R_4}, \alpha_4})\\
    &[h_{\alpha_3\alpha_4}(\boldsymbol{R_3-R_4}) \mathcal{I}_{\alpha_1\alpha_2\alpha_3\alpha_4}(\boldsymbol{R_1, R_2, R_3, R_4})
    + V_{\alpha_3\alpha_4}(\boldsymbol{R_3-R_4}) \mathcal{J}_{\alpha_1\alpha_2\alpha_3\alpha_4} (\boldsymbol{R_1, R_2, R_3, R_4})]
\end{aligned}
\label{eq:OM}
\end{equation}
\end{widetext}

\cref{eq:OM} is our central result. It shows that the OM of an interacting system depends only on exact zero-frequency response functions evaluated at zero magnetic field, which is not unexpected since a uniform magnetic field can be regarded as the zero-frequency limit of an electromagnetic wave.

\textbf{\textit{Locality condition.}}
\cref{eq:OM} also provides quantitative insight into the locality condition, which until now has only been heuristically assumed in Ref.~\cite{Shi2007QuantumOM} and in our derivation above.
Crucially, for the expression to be well defined in a periodic system, the long-range contributions to the summations $\sum_{1234} (x_1+x_2-x_3-x_4)\mathcal{I}(1,2,3,4)$ and $\sum_{1234} (x_1+x_2-x_3-x_4)\mathcal{J}(1,2,3,4)$ must converge.
Since we have assumed $h_{\alpha\beta}$ and $V_{\alpha\beta}$ to be exponentially short-ranged, we always have $\boldsymbol{R}_1\approx \boldsymbol{R}_2$ and $\boldsymbol{R}_3\approx \boldsymbol{R}_4$.
Thus, the convergence criteria is that $\mathcal{I}(1,2,3,4)$ and $\mathcal{J}(1,2,3,4)$ must decay faster than $|R^C_{12} - R^C_{34}|^{-2}$.
This justifies applying \cref{eq:OM} to a wide range of systems, including but not limited to insulating states and Fermi liquid states at both zero and finite temperatures and in two and three spatial dimensions.
Conversely, the failure of convergence in states with more extended response functions signals non-locality of the grand potential density, undermining the identification of the local density response with the OM in a uniform magnetic field.

It is worth noting that \cref{eq:OM} is not manifestly antisymmetric under $x\leftrightarrow y$.
Physically, this reflects that we assumed locality only along the $x$ direction in the derivation.
For a qualitatively isotropic system, one could instead take $\boldsymbol{q} = q \boldsymbol{\hat{y}}$ and $\boldsymbol{A}(\boldsymbol{r}) = -A_0 \sin (\boldsymbol{q\cdot r}) \boldsymbol{\hat{x}}$, and rederive $\tilde{M}_z$ by taking the long wavelength limit.
Averaging the two procedures would yield an explicitly antisymmetric expression.
As will be shown in the following examples, such a manual anti-symmetrization is unnecessary, since \cref{eq:OM} already yields the correct $\tilde{M}_z$.

\textbf{\textit{Non-interacting electrons.}}
For non-interacting electrons at finite temperature $T$, it is straightforward to write
\begin{equation}
\footnotesize
\begin{aligned}
    \mathcal{I}_{\alpha_1\alpha_2\alpha_3\alpha_4}(\boldsymbol{R_1, R_2, R_3, R_4}) =\frac 1{N^2} \sum_{\boldsymbol{k_1k_2}} C^{\mathcal{P}}_{c_{\boldsymbol{k_1},\alpha_1}^\dagger c_{\boldsymbol{k_2},\alpha_2},c_{\boldsymbol{k_2},\alpha_3}^\dagger c_{\boldsymbol{k_1},\alpha_4}} (\omega=0)\\
    \exp(-\ii \boldsymbol{k_1}\cdot \boldsymbol{\Delta R}_{14, \alpha_1 \alpha_4} +\ii \boldsymbol{k_2}\cdot \boldsymbol{\Delta R}_{23, \alpha_2\alpha_3})
\end{aligned}
\label{eq:Fourier_non_interacting}
\end{equation}
where we have denote $\boldsymbol{\Delta R}_{ij, \alpha_i\alpha_j} = \boldsymbol{R_i} + \boldsymbol{r}_{\alpha_i}-\boldsymbol{R_j} - \boldsymbol{r}_{\alpha_j}$.
Diagrammatically we have
\begin{widetext}
\begin{equation}
\small
\begin{aligned}
    C^{\mathcal{P}}_{c_{\boldsymbol{k_1},\alpha_1}^\dagger c_{\boldsymbol{k_2},\alpha_2},c_{\boldsymbol{k_2},\alpha_3}^\dagger c_{\boldsymbol{k_1},\alpha_4}} (\omega = 0)&= \sum_{m_1m_2} U_{\alpha_1m_1}^*(\boldsymbol{k_1}) U_{\alpha_2m_2}(\boldsymbol{k_2}) U_{\alpha_3m_2}^*(\boldsymbol{k_2}) U_{\alpha_4m_1}(\boldsymbol{k_1}) \frac{1}{\beta}\sum_{\omega_n} \frac{1}{(\ii \omega_n-\epsilon_{m_1}(\boldsymbol{k_1}))(\ii \omega_n-\epsilon_{m_2}(\boldsymbol{k_2}))}\\
    &=\sum_{m_1m_2} U_{\alpha_1m_1}^*(\boldsymbol{k_1}) U_{\alpha_2m_2}(\boldsymbol{k_2}) U_{\alpha_3m_2}^*(\boldsymbol{k_2}) U_{\alpha_4m_1}(\boldsymbol{k_1})  \mathcal{P}[\frac{f(\epsilon_{m_1}(\boldsymbol{k_1}))- f(\epsilon_{m_2}(\boldsymbol{k_2}))}{\epsilon_{m_1}(\boldsymbol{k_1}) -  \epsilon_{m_2}(\boldsymbol{k_2})}]
\end{aligned}
\label{eq:momentum_non_interacting}
\end{equation}
where $\omega_n$'s are the fermion Matsubara frequencies, $\epsilon _m(\boldsymbol{k}),U_{\alpha m}(\boldsymbol{k})$ are the single-particle energies and wavefunctions for the $m$th band, and $f(\epsilon)$ is the Fermi distribution.
The principle value part should be understood as $\partial f/\partial \epsilon|_{\epsilon_{m_1}(\boldsymbol{k_1})}$ if $\epsilon_{m_1}(\boldsymbol{k_1}) = \epsilon_{m_2}(\boldsymbol{k_2}) $.
Substituting \cref{eq:momentum_non_interacting} into \cref{eq:Fourier_non_interacting} and then \cref{eq:OM}, we find (see Ref.~\cite{sup} for details)
\begin{equation}
\small
\tilde{M}_z = -\frac{\ii e}{2\hbar} \sum_{\boldsymbol{k}m} \{(\langle \partial_x u_{m}(\boldsymbol{k})|\tilde{h}(\boldsymbol{k})+\epsilon_{m}(\boldsymbol{k})|\partial_y u_{m}(\boldsymbol{k})\rangle) f_{m}(\boldsymbol{k})- (\langle \partial_x u_m(\boldsymbol{k})|\epsilon_m(\boldsymbol{k})-\tilde{h}({\boldsymbol{k}})|\partial_y u_m(k)\rangle)\epsilon_m(\boldsymbol{k})[\partial f/\partial\epsilon]|_{\epsilon=\epsilon_m(k)}\} - (x\leftrightarrow y)
\label{eq:non_int_res}
\end{equation}
\end{widetext}
where we have denoted $\partial_{k_i} =\partial_i$, and $\tilde{h}(\boldsymbol{k})$ is the non-interacting Hamiltonian in momentum space. \cref{eq:non_int_res} is in perfect agreement with the auxiliary OM derived in Ref.~\cite{Shi2007QuantumOM}, with the understanding that we have absorbed the chemical potential $\mu$ into the definition of $\epsilon_m(\boldsymbol{k})$.
Importantly, while we started with \cref{eq:OM} which is not explicitly antisymmetric, the final expression automatically restores antisymmetricity, manifesting the gauge-invariance of the magnetic field.

\textbf{\textit{Weak coupling expansion to the first order.}}
In the weak-coupling limit, $\mathcal{I}$ and $\mathcal{J}$, and hence $\tilde{M}_z$, can be expanded order by order in the interaction strength.
We now calculate the derivative of $\tilde{M}_z$ to the interaction strength $U$ at fixed chemical potential and the weak-coupling limit, $k_U =\partial \tilde{M}_z/{\partial U}|_{U=0, \mu}$, which requires calculating first-order diagrams for $\mathcal{I}$ and zeroth-order diagrams for $\mathcal{J}$.
The resulting linear interaction correction to $\tilde{M}_z$ can be decomposed as $\delta \tilde{M}_z = \delta \tilde{M}_z^{(1)} + \delta \tilde{M}_z^{(2)} + \delta \tilde{M}_z^{(3)}$, where $\delta \tilde{M}_z^{(1)}$ and $\delta \tilde{M}_z^{(2)}$ originate from the self-energy and bare vertex corrections to $\mathcal{I}$, respectively, and $\delta \tilde{M}_z^{(3)}$ comes from the contribution of $\mathcal{J}$.
The detailed derivation of $\delta \tilde{M}_z^{(i)}$ and $k_U$ in terms of $V_{\alpha\beta}$ and the non-interacting band structure can be found in Ref.~\cite{sup}.

It was recently proven that the modern theory of OM can be generalized to self-consistent Hartree-Fock bands \cite{Kang2025HFOM,Liu2025TMDHFOM,ZhuHuang2025Projectors}.
In other words, \cref{eq:non_int_res} remains valid at the static mean-field level, provided that one replaces the non-interacting dispersions and wavefunctions with the self-consistent Hartree-Fock ones.
Since the Hartree-Fock approximation correctly captures first-order interaction effects in the weak-coupling limit, we expect $k_U$ to equal
$k_{HF} = \frac{\partial \tilde{M}^{HF}_z}{\partial U}|_{U=0, \mu}$,
where $\tilde{M}^{HF}_z(\mu, U)$ is the auxiliary OM from self-consistent Hartree-Fock calculations at fixed chemical potential $\mu$ and interaction strength $U$.

\begin{figure}[t!]
	\centering
    \includegraphics[width=1\linewidth]{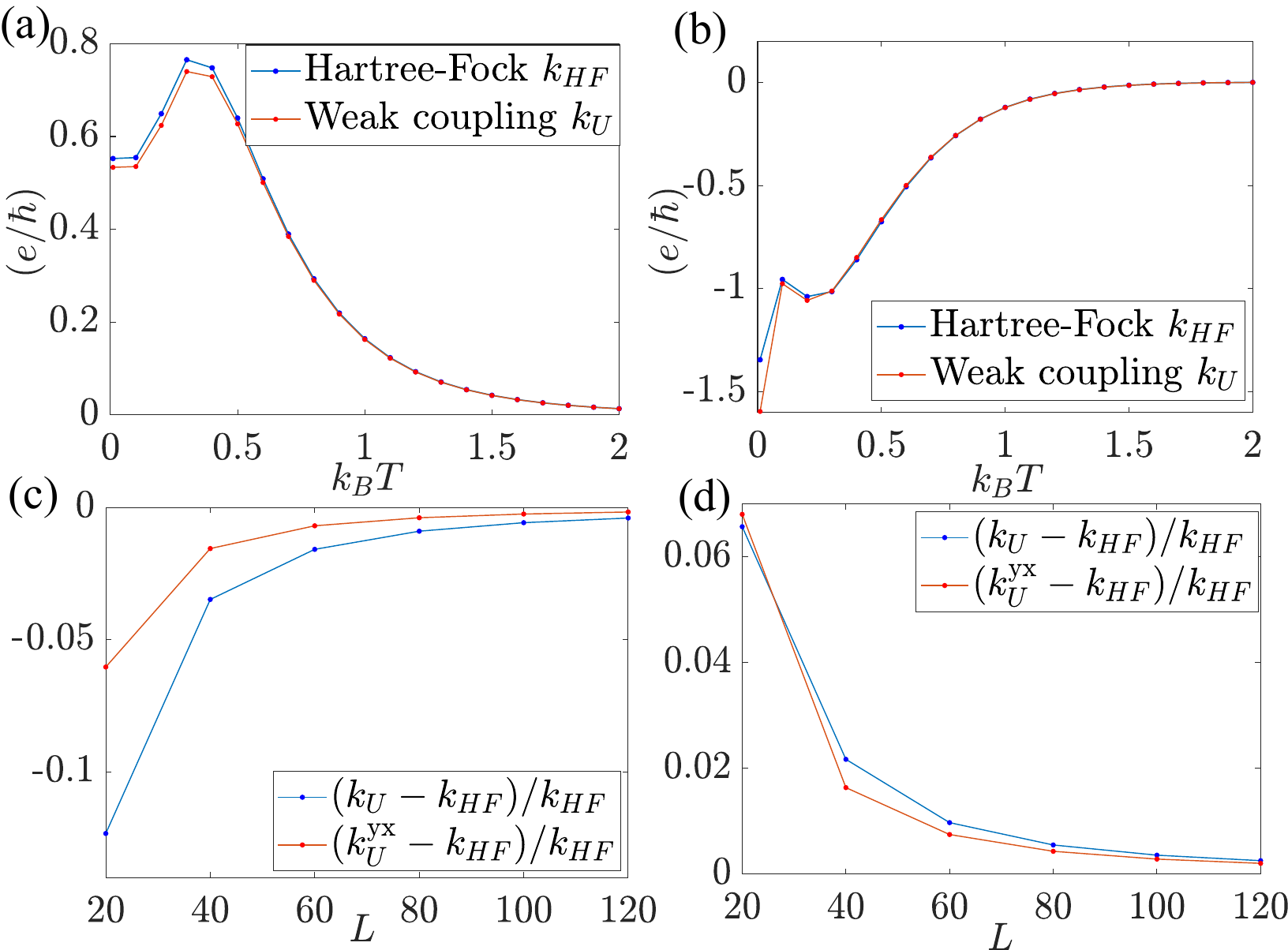}
    \caption{Comparison between the first order derivative of auxiliary OM $\tilde{M}_z$ to interaction strength $U$ calculated from Hartree-Fock bands ($k_{HF}$) and the weak-coupling expansion ($k_U$), as well as the result from permuting $x\leftrightarrow y$ in \cref{eq:OM} plus an additional minus sign ($k_U^{\rm yx}$).
    Panels (a) and (b) show the agreement between $k_U$ and $k_{HF}$ at system size $L=40$ for a Chern insulator ($\mu = 0$, $C=-1$) and a metal ($\mu = 1.5$), respectively.
    We deliberately choose a relatively small $L$ to make the two lines distinguishable to the eye.
    The small difference is due to finite-size effect.
    Panels (c) and (d) display the finite-size scaling of relative differences at low temperature ($k_BT=0.1$) corresponding to (a) and (b) respectively, plotting the relative difference between $k_{HF}$ and $k_U$, as well as between $k_{HF}$ and $k_{U}^{\rm yx}$ against $L$.
    The data confirm that both differences vanish in the thermodynamic limit.}
\label{fig_main1}
\end{figure}

To numerically confirm $k_U = k_{HF}$, we study a QWZ-like~\cite{PhysRevB.74.085308} spinless two-band model with unity lattice constant and the non-interacting Hamiltonian
\begin{equation}
\small
 \tilde{h}(\boldsymbol{k}) = t_1 \sin k_x \sigma_x + t_2 \sin k_y\sigma_y +  (m+\cos k_x + \cos k_y) \sigma_z - \mu \sigma_0
\end{equation}
where $\sigma_{x,y,z}$ are Pauli matrices in orbital space, $\sigma_0$ is the identity matrix, $m$ is the mass term, and $\mu$ is the chemical potential. The two orbitals in the same unit cell are assumed to be at the same position. The interaction is included as an extended Hubbard-like one with orbital dependence,
$\tilde{V}_{\alpha\beta}(\boldsymbol{p}) = U_0(\sigma_0 + w_0 \sigma_x)_{\alpha\beta} (1+\cos p_x + \cos p_y)$.
Without loss of generality, we fix $t_1=1.5, t_2= m=1.0, w_0=0.5$, and compare $k_{HF}$ and $k_U$ at different $\mu$ and temperature $T$.

As illustrated in \cref{fig_main1}~(a) and (b), $k_{HF}$ and $k_U$ already show good agreement at a moderate system size of $L = 40$, both for the Chern insulator state at $\mu = 0$ ($C=-1$) and the metallic state at $\mu = 1.5$. This agreement is particularly strong at higher temperatures.
To rigorously test the convergence at low temperature, we analyze the scaling of relative differences with linear system size $L$ at $k_BT=0.1$ in \cref{fig_main1}~(c) and (d), where we plot the relative difference between $k_{HF}$ and $k_{U}$, alongside the difference between $k_{HF}$ and $k_U^{\rm yx}$, which is defined by exchanging $x \leftrightarrow y$ in \cref{eq:OM} plus an additional minus sign, against linear system size $L$.
We observe that both differences decay rapidly towards zero as $L$ increases. This confirms that $k_U$ converges to $k_{HF}$ and that the $x\leftrightarrow y$ antisymmetry is automatically restored in the thermodynamic limit, thereby validating \cref{eq:OM} to first order in the weak-coupling expansion.

\textbf{\textit{Discussion.}}
In summary, we propose an exact formalism for the OM of electrons with short-range interactions invariant under Peirels substitution based on the locality of grand potential density response.
While we benchmarked the approach using a weak-coupling expansion, the derivation of \cref{eq:OM} is not perturbative in the interaction strength and is therefore expected to remain valid even for strongly correlated states far from the non-interacting limit.
While we used a tight-binding model in the derivation, the formula is readily applicable to realistic materials, including moir\'e systems, by noting that there are many orbitals within each moir\'e unit cell.
\cref{eq:OM} can also be easily generalized to continuum models by replacing the composite label $\{\boldsymbol{R}, \alpha\} \to \boldsymbol{r}$ and $\sum_{\boldsymbol{R}, \alpha} \to \int d\boldsymbol{r}$.
This formula bridges experimentally relevant magnetic phenomena and standard theoretical many-body techniques, with a broad range of applicability.
The framework can be combined with numerical and analytical many-body methods that directly access response functions, offering a route towards genuinely strongly interacting yet controlled approximations for the quantitative evaluation of OM beyond mean-field treatments.

\textbf{\textit{Notes added.}} Upon concluding this work, we become aware of a very recent study~\cite{ye2026quantummanybodyapproachorbital}, which also derives an exact formula for OM of interacting electrons based on the Luttinger-Ward functional and using noncommutative coordinates.
In contrast, the present work focuses on short-ranged interactions and employs the locality of grand potential density to express OM only in terms of zero frequency response functions, which are directly accessible to many many-body techniques.

\textbf{\textit{Acknowledgments.}}
Z.-D.~S. and X.~C. were supported by National Natural Science Foundation of China (General Program No.~12274005), National Key Research and Development Program of China (No.~2021YFA1401900), and Quantum Science and Technology-National Science and Technology Major Project (No.~2021ZD0302403).

\clearpage
\onecolumngrid 
\appendix{\bf{Appendix}}

\crefname{appendix}{Sec.}{Secs.}
\crefname{equation}{Eq.}{Eqs.}
\crefname{figure}{Fig.}{Figs.}
\crefname{table}{Table}{Tables}
\crefname{section}{Sec.}{Secs.}

\renewcommand{\paragraph}[1]{\vspace{0.2cm}{\bf \textit{#1}}}
\renewcommand\thesection{\Roman{section}}
\renewcommand\thesubsection{\Alph{subsection}}
\def\ie{i.e.,\ }
\def\eg{e.g.,\ }
\def\etc{etc.\ }

\definecolor{Gray}{gray}{0.85}
\newcolumntype{a}{>{\columncolor{Gray}}c}

\allowdisplaybreaks[1] %

\def\pare#1{\left( #1 \right)}
\def\brak#1{\left[#1\right]}
\def\brace#1{\left\{#1\right\}}
\def\bra#1{\langle #1 |}
\def\ket#1{| #1 \rangle}
\def\Bra#1{\left\langle #1 \right|}
\def\Ket#1{\left| #1 \right\rangle}
\def\inn#1{\langle #1 \rangle}
\def\Inn#1{\left\langle #1 \right\rangle}
\def\abs#1{\left| #1 \right|}
\def\Im{\mathrm{Im}}
\def\Re{\mathrm{Re}}

\def\mF{\mathcal{F}}
\def\ii{\mathrm{i}}
\def\pp{\mathbf{p}}
\def\Tr{\mathrm{Tr}}
\def\kk{\mathbf{k}}
\def\GG{\mathbf{G}}
\def\mA{\mathcal{A}}
\def\mD{\mathcal{D}}
\def\RR{\mathbf{R}}
\def\tt{\mathbf{t}}

\renewcommand{\thefigure}{S\arabic{figure}}
\renewcommand{\theequation}{S\arabic{equation}}
\setcounter{figure}{0}

\section{Calculation of $\tilde{M}_z$ for non-interacting electrons}
In this section we show in detail the derivation from  \cref{eq:momentum_non_interacting} to \cref{eq:non_int_res} of the main text.
The non-interacting Hamiltonian in momentum space is
\begin{equation}
    \tilde{h}_{\alpha\beta}(\boldsymbol{k}) = \sum_{\boldsymbol{\Delta R}} h_{\alpha\beta}(\boldsymbol{\Delta R}) e^{-\ii \boldsymbol{k\cdot (\Delta R+ r_{\alpha\beta})}}
\end{equation}
The non-interacting energy bands and wavefunctions $\epsilon_m(\boldsymbol{k})$ and $U_{\alpha m}(\boldsymbol{k})$ are eigenvalues and eigenstates of $\tilde{h}_{\alpha\beta}(\boldsymbol{k})$, labeled by the band index $m$. 
Substituting \cref{eq:momentum_non_interacting} to \cref{eq:Fourier_non_interacting} and then \cref{eq:OM} of the main text and using integration-by-parts, we have
\begin{equation}
\footnotesize
\begin{aligned}
    \tilde{M}_z &= \frac{\ii e}{2\hbar N^2} \sum_{\boldsymbol{R_1...R_4}} \sum_{\alpha_1...\alpha_4} \sum_{m_1m_2} \sum_{\boldsymbol{k_1k_2}}(y_{\boldsymbol{R_1}, \alpha_1} - y_{\boldsymbol{R_2}, \alpha_2})(x_{\boldsymbol{R_1}, \alpha_1} + x_{\boldsymbol{R_2}, \alpha_2} - x_{\boldsymbol{R_3, \alpha_3}}- x_{\boldsymbol{R_4}, \alpha_4}) h_{\alpha_1\alpha_2}(\boldsymbol{R_1-R_2}) h_{\alpha_3\alpha_4}(\boldsymbol{R_3-R_4})\\
    & \qquad\qquad\qquad  \qquad\qquad\qquad e^{-\ii \boldsymbol{k_1\cdot R}_{14, \alpha_1\alpha_4}} e^{\ii \boldsymbol{k_2\cdot R}_{23, \alpha_2\alpha_3}}U_{\alpha_1m_1}^*(\boldsymbol{k_1}) U_{\alpha_2m_2}(\boldsymbol{k_2}) U_{\alpha_3m_2}^*(\boldsymbol{k_2}) U_{\alpha_4m_1}(\boldsymbol{k_1}) \mathcal{P}[\frac{f(\epsilon_{m_1}(\boldsymbol{k_1})) - f(\epsilon_{m_2}(\boldsymbol{k_2}))}{\epsilon_{m_1}(\boldsymbol{k_1}) - \epsilon_{m_2}(\boldsymbol{k_2})}]\\
    & = \frac{\ii e}{2\hbar N^4} \sum_{\boldsymbol{R_1...R_4}}\sum_{\alpha_1...\alpha_4} \sum_{m_1m_2} \sum_{\boldsymbol{k_1...k_4}} (y_{\boldsymbol{R_1}, \alpha_1} - y_{\boldsymbol{R_2}, \alpha_2})(x_{\boldsymbol{R_1}, \alpha_1} + x_{\boldsymbol{R_2}, \alpha_2} - x_{\boldsymbol{R_3, \alpha_3}}- x_{\boldsymbol{R_4}, \alpha_4}) h_{\alpha_1\alpha_2}(\boldsymbol{k_3}) h_{\alpha_3\alpha_4}(\boldsymbol{k_4})\\
    & \qquad e^{-\ii \boldsymbol{k_1\cdot R}_{14, \alpha_1\alpha_4}} e^{\ii \boldsymbol{k_2\cdot R}_{23, \alpha_2\alpha_3}} e^{\ii \boldsymbol{k_3\cdot R}_{12, \alpha_1\alpha_2}} e^{\ii \boldsymbol{k_4\cdot R}_{34, \alpha_3\alpha_4}} U_{\alpha_1m_1}^*(\boldsymbol{k_1}) U_{\alpha_2m_2}(\boldsymbol{k_2}) U_{\alpha_3m_2}^*(\boldsymbol{k_2}) U_{\alpha_4m_1}(\boldsymbol{k_1}) \mathcal{P}[\frac{f(\epsilon_{m_1}(\boldsymbol{k_1})) - f(\epsilon_{m_2}(\boldsymbol{k_2}))}{\epsilon_{m_1}(\boldsymbol{k_1}) - \epsilon_{m_2}(\boldsymbol{k_2})}]\\
    &=\frac{\ii e}{2\hbar} \sum_{\alpha_1...\alpha_4}\sum_{\boldsymbol{k}} \sum_{m_1m_2} \{\partial_y \tilde{h}_{\alpha_1\alpha_2}(\boldsymbol{k}) \tilde{h}_{\alpha_3\alpha_4}(\boldsymbol{k})(\partial_{k_1^x} - \partial_{k_2^x}) \{ U_{\alpha_1m_1}^*(\boldsymbol{k_1}) U_{\alpha_2m_2}(\boldsymbol{k_2}) U_{\alpha_3m_2}^*(\boldsymbol{k_2}) U_{\alpha_4m_1}(\boldsymbol{k_1}) \mathcal{P}[\frac{f(\epsilon_{m_1}(\boldsymbol{k_1})) - f(\epsilon_{m_2}(\boldsymbol{k_2}))}{\epsilon_{m_1}(\boldsymbol{k_1}) - \epsilon_{m_2}(\boldsymbol{k_2})}]\}|_{\boldsymbol{k_1=k_2=k}}\\
    &=\frac{\ii e}{2\hbar} \sum_{\boldsymbol{k}} \{\sum_{mn} (\langle \partial_x u_{m}(\boldsymbol{k})|u_n(\boldsymbol{k})\rangle \langle u_n(\boldsymbol{k})|\partial_y\tilde{h}(\boldsymbol{k})|u_{m}(\boldsymbol{k})\rangle - \langle u_{m}(\boldsymbol{k})|\partial_y \tilde{h}(\boldsymbol{k})|u_n(\boldsymbol{k})\rangle \langle u_n(\boldsymbol{k})|\partial_x u_{m}(\boldsymbol{k})\rangle) \epsilon_m(\boldsymbol{k}) f'(\epsilon_m(\boldsymbol{k}))\\&\qquad\qquad\qquad+\sum_{m_1m_2}\langle u_{m_1}(\boldsymbol{k})|\partial_y\tilde{h}(\boldsymbol{k})| u_{m_2}(\boldsymbol{k})\rangle (\epsilon_{m_1}(\boldsymbol{k}) + \epsilon_{m_2}(\boldsymbol{k})) \langle u_{m_2}(\boldsymbol{k})|\partial_x u_{m_1}(\boldsymbol{k})\rangle \mathcal{P}[\frac{f(\epsilon_{m_1}(\boldsymbol{k})) - f(\epsilon_{m_2}(\boldsymbol{k}))}{\epsilon_{m_1}(\boldsymbol{k}) - \epsilon_{m_2}(\boldsymbol{k})}]\}
\end{aligned}
\label{eq:MZ_free_s1}
\end{equation}
Using $\langle u_m(\boldsymbol{k})|\partial_\mu\tilde{h}(\boldsymbol{k})|u_n(\boldsymbol{k})\rangle = \delta_{mn}\partial_\mu \epsilon_m(\boldsymbol{k}) + (\epsilon_m(\boldsymbol{k} ) - \epsilon_n(\boldsymbol{k}))]\langle \partial_x u_{m}(\boldsymbol{k})|u_{n}(\boldsymbol{k})\rangle$, \cref{eq:MZ_free_s1} can be simplified to
\begin{equation}
\footnotesize
\begin{aligned}
    \tilde{M}_z &= \frac{\ii e}{2\hbar} \sum_{\boldsymbol{k}, mn} \epsilon_m(\boldsymbol{k}) f'(\epsilon_m(\boldsymbol{k})) (\epsilon_m(\boldsymbol{k}) - \epsilon_n(\boldsymbol{k})) \{ \langle \partial_x u_m(\boldsymbol{k})| u_n(\boldsymbol{k})\rangle \langle u_n(\boldsymbol{k})|\partial_y u_m(\boldsymbol{k})\rangle - (x\leftrightarrow y)\}\\
    &+\frac{\ii e}{2\hbar} \sum_{\boldsymbol{k}, m_1m_2} \langle \partial_y u_{m_1}(\boldsymbol{k})| u_{m_2}(\boldsymbol{k})\rangle (\epsilon_{m_1}(\boldsymbol{k}) + \epsilon_{m_2}(\boldsymbol{k}))\langle u_{m_2}(\boldsymbol{k})| \partial_x u_{m_1}(\boldsymbol{k})\rangle [f(\epsilon_{m_1}(\boldsymbol{k})) - f(\epsilon_{m_2}(\boldsymbol{k}))]\\
    &=-\frac{\ii e}{2\hbar} \sum_{\boldsymbol{k}m} \{(\langle \partial_x u_{m}(\boldsymbol{k})|\tilde{h}(\boldsymbol{k})+\epsilon_{m}(\boldsymbol{k})|\partial_y u_{m}(\boldsymbol{k})\rangle) f_{m}(\boldsymbol{k})- (\langle \partial_x u_m(\boldsymbol{k})|\epsilon_m(\boldsymbol{k})-\tilde{h}({\boldsymbol{k}})|\partial_y u_m(k)\rangle)\epsilon_m(\boldsymbol{k})[\partial f/\partial\epsilon]|_{\epsilon=\epsilon_m(k)}\} - (x\leftrightarrow y)
\end{aligned}
\end{equation}
where we have recovered \cref{eq:non_int_res} of the main text.

\section{Diagrammatic calculation of $k_U =\partial \tilde{M}_z/{\partial U}|_{U=0, \mu}$}
\begin{figure}[t!]
	\centering
    \includegraphics[width=1\linewidth]{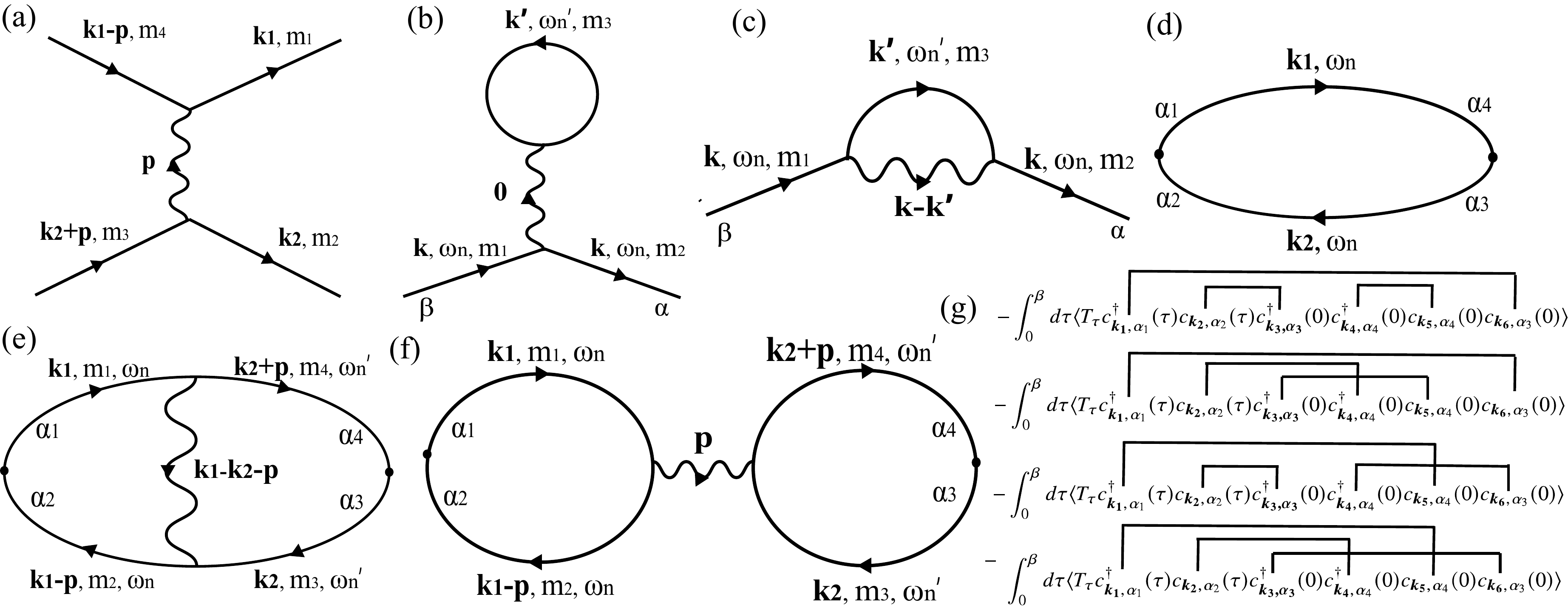}
    \caption{(a) A standard bare interaction vertex. (b-c) Hartree and Fock contributions to first order self-energy correction of the propagator. (d) The zeroth-order diagram for $\mathcal{I}$. First order self-energy correction to $\mathcal{I}$ should be dressing (b-c) to the two fermion lines in this diagram. (e-f) Bare vertex contribution to the first order result of $\mathcal{I}$. (g) Four different connected contractions for the zeroth order result of $\mathcal{J}$.}
\label{fig_sm1}
\end{figure}

In this section, we calculate $k_U =\partial \tilde{M}_z/{\partial U}|_{U=0, \mu}$ by a diagrammatic expansion of $\mathcal{I}_{\alpha_1\alpha_2\alpha_3\alpha_4}(\boldsymbol{R_1}, \boldsymbol{R_2}, \boldsymbol{R_3},\boldsymbol{R_4})$ and $\mathcal{J}_{\alpha_1\alpha_2\alpha_3\alpha_4}(\boldsymbol{R_1}, \boldsymbol{R_2}, \boldsymbol{R_3},\boldsymbol{R_4})$.
According to \cref{eq:OM} of the main text we have to calculate the first order diagrams for $\mathcal{I}$ and zeroth order diagrams for $\mathcal{J}$.
Using the Fourier transformations
\begin{equation}
    c_{\boldsymbol{R}, \alpha}^\dagger = \frac{1}{\sqrt{N}} \sum_{\boldsymbol{k}}e^{-\ii \boldsymbol{k\cdot R_\alpha}} c_{\boldsymbol{k}, \alpha}^\dagger, \qquad c_{\boldsymbol{R}, \alpha} = \frac{1}{\sqrt{N}} \sum_{\boldsymbol{k}}e^{\ii \boldsymbol{k\cdot R_{\alpha}}} c_{\boldsymbol{k}, \alpha}
\label{eq:operator_FT}
\end{equation}
where $\boldsymbol{R_\alpha = R+r_\alpha}$ and $N$ is the number of unit cells, we have
\begin{equation}
\begin{aligned}
    \mathcal{I}_{\alpha_1\alpha_2\alpha_3\alpha_4}(\boldsymbol{R_1}, \boldsymbol{R_2}, \boldsymbol{R_3}, \boldsymbol{R_4}) &= \frac{1}{N^2} \sum_{\boldsymbol{k_1}\boldsymbol{k_2}\boldsymbol{k_3}\boldsymbol{k_4}} e^{-\ii \boldsymbol{k_1\cdot R}_{1,\alpha_1}+\ii \boldsymbol{k_2\cdot R}_{2,\alpha_2}-\ii \boldsymbol{k_3\cdot R}_{3,\alpha_3}+\ii \boldsymbol{k_4\cdot R}_{4,\alpha_4}} C^{\mathcal{P}}_{c_{\boldsymbol{k_1},\alpha_1}^\dagger c_{\boldsymbol{k_2}, \alpha_2}; c_{\boldsymbol{k_3},\alpha_3}^\dagger c_{\boldsymbol{k_4}, \alpha_4}}(\omega = 0)\\
    & = \frac{1}{N^2} \sum_{\boldsymbol{k_1k_2p}} e^{-\ii \boldsymbol{k_1\cdot R}_{12, \alpha_1\alpha_2} - \ii \boldsymbol{k_2\cdot R}_{34, \alpha_3\alpha_4}} e^{-\ii \boldsymbol{p\cdot R}_{24, \alpha_2\alpha_4}} C^{\mathcal{P}}_{c_{\boldsymbol{k_1},\alpha_1}^\dagger c_{\boldsymbol{k_1-p}, \alpha_2}; c_{\boldsymbol{k_2},\alpha_3}^\dagger c_{\boldsymbol{k_2+p}, \alpha_4}}(\omega = 0)
\end{aligned}
\label{eq:I_express}
\end{equation}
where we have made use of momentum conservation in the second line, and
\begin{equation}
\begin{aligned}
    \mathcal{J}_{\alpha_1\alpha_2\alpha_3\alpha_4}(\boldsymbol{R_1}, \boldsymbol{R_2}, \boldsymbol{R_3}, \boldsymbol{R_4}) = \frac{1}{N^3} \sum_{\boldsymbol{k_1}...\boldsymbol{k_6}} e^{-\ii \boldsymbol{k_1\cdot R}_{1,\alpha_1} + \ii \boldsymbol{k_2\cdot R}_{2,\alpha_2} - \ii \boldsymbol{k_3\cdot R}_{3,\alpha_3} -\ii \boldsymbol{k_4\cdot R}_{4,\alpha_4} + \ii \boldsymbol{k_5\cdot R}_{4,\alpha_4} +\ii \boldsymbol{k_6\cdot R}_{3,\alpha_3}} \\C^{\mathcal{P}}_{c_{\boldsymbol{k_1},\alpha_1}^\dagger c_{\boldsymbol{k_2}, \alpha_2}; c_{\boldsymbol{k_3},\alpha_3}^\dagger c_{\boldsymbol{k_4}, \alpha_4}^\dagger c_{\boldsymbol{k_5}, \alpha_4} c_{\boldsymbol{k_6}, \alpha_3}}(\omega = 0)
\end{aligned}
\label{eq:J_express}
\end{equation}
We now specify diagrammatic rules of the weak coupling expansion. Substituting \cref{eq:operator_FT} into \cref{eq:Hamiltonian} of the main text, we have the many-body Hamiltonian in momentum space
\begin{equation}
    \hat{K}_0 = \sum_{\alpha\beta}\sum_{\boldsymbol{k}} \tilde{h}_{\alpha\beta}(\boldsymbol{k}) c_{\boldsymbol{k}, \alpha}^\dagger c_{\boldsymbol{k},\beta} + \frac{1}{N} \sum_{\boldsymbol{k_1k_2p}}\sum_{\alpha\beta} \tilde{V}_{\alpha\beta}(\boldsymbol{p}) c_{\boldsymbol{k_1}, \alpha}^\dagger c_{\boldsymbol{k_2}, \beta}^\dagger c_{\boldsymbol{k_2+p}, \beta} c_{\boldsymbol{k_1-p}, \alpha}
\label{eq:K0_orbitalbasis}
\end{equation}
where $\tilde{V}_{\alpha\beta}(\boldsymbol{p}) = \sum_{\boldsymbol{\Delta R}} V_{\alpha\beta}(\boldsymbol{\Delta R}) e^{-\ii \boldsymbol{p\cdot (\Delta R+ r_{\alpha\beta})}}$.
Transforming \cref{eq:K0_orbitalbasis} to the band basis yields
\begin{equation}
    \hat{K}_0 = \sum_{\boldsymbol{k} ,m} \epsilon_m(\boldsymbol{k}) c_{\boldsymbol{k}, m}^\dagger c_{\boldsymbol{k}, m} + \frac{1}{N} \sum_{\boldsymbol{k_1k_2p}} \sum_{m_1m_2m_3m_4} V_{m_1m_2m_3m_4}(\boldsymbol{k_1}, \boldsymbol{k_2}, \boldsymbol{p}) c_{\boldsymbol{k_1}, m}^\dagger c_{\boldsymbol{k_2}, m_2}^\dagger c_{\boldsymbol{k_2+p}, m_3} c_{\boldsymbol{k_1-p}, m_4}
\label{eq:K0_bandbasis}
\end{equation}
where $c_{\boldsymbol{k}, m}^\dagger = \sum_{\alpha} U_{\alpha m}(\boldsymbol{k}) c_{\boldsymbol{k}, \alpha}^\dagger$ is the Fermion creation operator in th $m$th non-interacting band, and
\begin{equation}
    V_{m_1m_2m_3m_4}(\boldsymbol{k_1, k_2, p}) = \sum_{\alpha\beta}\tilde{V}_{\alpha\beta}(\boldsymbol{p}) U_{\alpha m_1}^*(\boldsymbol{k_1}) U_{\beta m_2}^*(\boldsymbol{k_2})U_{\beta m_3}(\boldsymbol{k_2+p}) U_{\alpha m_4}(\boldsymbol{k_1-p})
\end{equation}
From \cref{eq:K0_bandbasis} we can read out the diagrammatic rules: assign $-\frac{1}{\ii \omega_n - \epsilon_m(\boldsymbol{k})}$ to each free propagator line labeled by band index $m$, momentum $\boldsymbol{k}$ and fermion Matsubara frequency $\omega_n$. For each bare interaction vertex illustrated in \cref{fig_sm1}~(a) assign a factor $-\frac{2}{N}V_{m_1m_2m_3m_4}(\boldsymbol{k_1}, \boldsymbol{k_2}, \boldsymbol{p})$, where the factor $2$ comes from two different ways to assign bare vertex legs to propagator lines.
For a propagator starting and ending at the same vertex, an $e^{\ii \omega_n\eta}$ factor with $\eta = 0^+$ should be assigned due to the normal-ordering of the interaction.
All internal labels of a diagram (including momenta, frequencies, and band indices) should be summed over.
Each closed fermion loop generates an additional $-1$ factor.
The diagrammatic result of each response function is further multiplied by another $-1$ due to our convention of response functions (see \cref{eq:Lehmann_rep_zerofreq} and Ref.~\cite{AltlandSimons2010CMFT}).

We first focus on \cref{eq:I_express}, including both self-energy and bare vertex contributions.
The first order self-energy contribution to a propagator includes both Hartree and Fock diagrams as illustrated in \cref{fig_sm1}~(b-c), summing to
\begin{equation}
\footnotesize
\begin{aligned}
    G^{(1)}_{\alpha\beta}(\boldsymbol{k},\ii \omega_n) &= \beta^{-1}\sum_{m_1m_2m_3} \sum_{\boldsymbol{k'}, \omega_n'}U_{\alpha m_2}(\boldsymbol{k})U_{\beta m_1}^*(\boldsymbol{k}) \frac{2}{N}( V_{m_2m_3m_3m_1}(\boldsymbol{k}, \boldsymbol{k'}, 0) - V_{m_2m_3m_1m_3}(\boldsymbol{k}, \boldsymbol{k'}, \boldsymbol{k-k'})) 
    \frac{-1}{\ii \omega_n - \epsilon_{m_1}(\boldsymbol{k})} \frac{-1}{\ii \omega_n - \epsilon_{m_2}(\boldsymbol{k})} \frac{-e^{\ii \omega_n'\eta}}{\ii \omega_n' - \epsilon_{m_3}(\boldsymbol{k'})}\\
    &=-\beta^{-1}\sum_{m_1m_2} U_{\alpha m_2}(\boldsymbol{k})U_{\beta m_1}^*(\boldsymbol{k}) S^{m_2m_1}(\boldsymbol{k}) \frac{1}{\ii \omega_n - \epsilon_{m_1}(\boldsymbol{k})} \frac{1}{\ii \omega_n - \epsilon_{m_2}(\boldsymbol{k})}
\end{aligned}
\end{equation}
where
\begin{equation}
    S^{m_2m_1}(\boldsymbol{k}) = \frac{2}{N}\sum_{m_3\boldsymbol{k'}} ( V_{m_2m_3m_3m_1}(\boldsymbol{k}, \boldsymbol{k'}, 0) - V_{m_2m_3m_1m_3}(\boldsymbol{k}, \boldsymbol{k'}, \boldsymbol{k-k'})) f(\epsilon_{m_3}(\boldsymbol{k'}))
\end{equation}
and we have made use of
\begin{equation}
    \beta^{-1}\sum_{\omega_n} \frac{e^{\ii \omega_n\eta}}{\ii \omega_n - \epsilon} =f(\epsilon)
\end{equation}
where $f(\epsilon)$ is the Fermi distribution.
Hence, the first-order self energy part of $C^{\mathcal{P}}_{c_{\boldsymbol{k_1}, \alpha_1}^\dagger c_{\boldsymbol{k_1-p}, \alpha_2}; c_{\boldsymbol{k_2}, \alpha_3}^\dagger c_{\boldsymbol{k_2+p},\alpha_4}} (\omega = 0)$, corresponding to dressing the first order self energy onto the two respective fermion lines in \cref{fig_sm1}~(d), sum to
\begin{equation}
\begin{aligned}
    &\delta_{\boldsymbol{p}, \boldsymbol{k_1-k_2}} \beta^{-1}\sum_{\omega_n} (G^{(1)}_{\alpha_4\alpha_1}(\boldsymbol{k_1}, \ii \omega_n) G^{(0)}_{\alpha_2\alpha_3}(\boldsymbol{k_2}, \ii \omega_n) + G^{(0)}_{\alpha_4\alpha_1}(\boldsymbol{k_1}, \ii \omega_n) G^{(1)}_{\alpha_2\alpha_3}(\boldsymbol{k_2}, \ii \omega_n))\\
    =& \delta_{\boldsymbol{p}, \boldsymbol{k_1-k_2}} \beta^{-1}\sum_{\omega_n} \sum_{m_1m_2m_3} \{ U_{\alpha_1m_1}^*(\boldsymbol{k_1}) U_{\alpha_2m_3}(\boldsymbol{k_2})U_{\alpha_3m_3}^*(\boldsymbol{k_2})U_{\alpha_4 m_2}(\boldsymbol{k_1}) \frac{1}{\ii \omega_n - \epsilon_{m_1}(\boldsymbol{k_1})} \frac{1}{\ii \omega_n -\epsilon_{m_2}(\boldsymbol{k_1})}\frac{1}{\ii \omega_n - \epsilon_{m_3}(\boldsymbol{k_2})} S^{m_2m_1}(\boldsymbol{k_1})\\
    &\qquad\qquad\qquad\qquad+U_{\alpha_1m_1}^*(\boldsymbol{k_1})U_{\alpha_2m_2}(\boldsymbol{k_2}) U_{\alpha_3m_3}^*(\boldsymbol{k_2}) U_{\alpha_4m_1}(\boldsymbol{k_1}) \frac{1}{\ii \omega_n -\epsilon_{m_1}(\boldsymbol{k_1})}\frac{1}{\ii \omega_n - \epsilon_{m_2}(\boldsymbol{k_2})}\frac{1}{\ii \omega_n - \epsilon_{m_3}(\boldsymbol{k_2})}S^{m_2m_3}(\boldsymbol{k_2})\}\\
    =& \delta_{\boldsymbol{p}, \boldsymbol{k_1-k_2} }\mathcal{A}_{\alpha_1\alpha_2\alpha_3\alpha_4}(\boldsymbol{k_1}, \boldsymbol{k_2})
\end{aligned}
\label{eq:self_energy_firstorder}
\end{equation}
where
\begin{equation}
\begin{aligned}
    \mathcal{A}_{\alpha_1\alpha_2\alpha_3\alpha_4}(\boldsymbol{k_1}, \boldsymbol{k_2}) &= \sum_{m_1m_2m_3} \{U_{\alpha_1m_1}^*(\boldsymbol{k_1}) U_{\alpha_2m_3}(\boldsymbol{k_2})U_{\alpha_3m_3}^*(\boldsymbol{k_2}) U_{\alpha_4m_2}(\boldsymbol{k_1}) D^{(3)}[\epsilon_{m_1}(\boldsymbol{k_1}), \epsilon_{m_2}(\boldsymbol{k_1}), \epsilon_{m_3}(\boldsymbol{k_2})] S^{m_2m_1}(\boldsymbol{k_1})\\
    &\qquad\qquad+ U_{\alpha_1m_1}^*(\boldsymbol{k_1}) U_{\alpha_2m_2}(\boldsymbol{k_2}) U_{\alpha_3m_3}^*(\boldsymbol{k_2}) U_{\alpha_4m_1}(\boldsymbol{k_1}) D^{(3)}[\epsilon_{m_1}(\boldsymbol{k_1}), \epsilon_{m_2}(\boldsymbol{k_2}), \epsilon_{m_3}(\boldsymbol{k_2})] S^{m_2m_3}(\boldsymbol{k_2})\}
\end{aligned}
\end{equation}
and we have denoted
\begin{equation}
    D^{(3)}[\epsilon_1, \epsilon_2, \epsilon_3] = \beta^{-1}\sum_{\omega_n}\prod_{i=1}^{3} \frac{1}{\ii \omega_n - \epsilon_i} = \sum_{i=1}^n \mathcal{P} [\frac{f(\epsilon_i)}{\prod_{j\neq i}(\epsilon_i - \epsilon_j)}]
\end{equation}
which is the third order divided difference of $\epsilon_1, \epsilon_2,\epsilon_3$. The principle value $\mathcal{P}$ means that if any singularity arises because $\epsilon_i = \epsilon_j = \epsilon$, we should take the limit $\epsilon_i = \epsilon+\eta, \epsilon_j = \epsilon-\eta$ and $\eta \to 0$ to avoid the singularity.
Specifically, if $\epsilon_1 = \epsilon_2 = \epsilon_3 = \epsilon$, we have
$D^{(3)}[\epsilon, \epsilon, \epsilon] = \frac{1}{2} f''(\epsilon)$.
Substituting \cref{eq:self_energy_firstorder} into \cref{eq:I_express} and then \cref{eq:OM} of the main text and using integration-by-parts, we have
\begin{equation}
\small
\begin{aligned}
    \delta \tilde{M}^{(1)}_{z} &= \frac{\ii e}{2\hbar N^4} \sum_{\boldsymbol{R_1}...\boldsymbol{R_4}} \sum_{\alpha_1...\alpha_4} \sum_{\boldsymbol{k_1k_2k_3k_4}} (y_{\boldsymbol{R}_1, \alpha_1} - y_{\boldsymbol{R}_2, \alpha_2})
    (x_{\boldsymbol{R_1}, \alpha_1} + x_{\boldsymbol{R_2}, \alpha_2} - x_{\boldsymbol{R_3}, \alpha_3} - x_{\boldsymbol{R_4}, \alpha_4}) \\&
    \qquad\qquad \qquad \qquad \qquad \qquad \qquad e^{-\ii \boldsymbol{k_1\cdot R}_{14, \alpha_1\alpha_4} + \ii \boldsymbol{k_2\cdot R}_{23, \alpha_2\alpha_3} + \ii \boldsymbol{k_3 \cdot R}_{12, \alpha_1\alpha_2} + \ii \boldsymbol{k_4\cdot R}_{34, \alpha_3\alpha_4}} \mathcal{A}_{\alpha_1\alpha_2\alpha_3\alpha_4}(\boldsymbol{k_1}, \boldsymbol{k_2}) \tilde{h}_{\alpha_1\alpha_2}(\boldsymbol{k_3}) \tilde{h}_{\alpha_3\alpha_4}(\boldsymbol{k_4})\\
    &=\frac{\ii e}{2\hbar N^4} \sum_{\boldsymbol{R_1...R_4}} \sum_{\alpha_1...\alpha_4} \sum_{\boldsymbol{k_1k_2k_3k_4}} \partial_{k_3^y} (\partial_{k_1^x} - \partial_{k_2^x}) [e^{-\ii \boldsymbol{k_1\cdot R}_{14, \alpha_1\alpha_4} + \ii \boldsymbol{k_2\cdot R}_{23, \alpha_2\alpha_3} - \ii \boldsymbol{k_3 \cdot R}_{12, \alpha_1\alpha_2} - \ii \boldsymbol{k_4\cdot R}_{34, \alpha_3\alpha_4}}]\\
    &\qquad\qquad\qquad \qquad\qquad\qquad\qquad\qquad\qquad\qquad\qquad\qquad\qquad\mathcal{A}_{\alpha_1\alpha_2\alpha_3\alpha_4}(\boldsymbol{k_1}, \boldsymbol{k_2}) \tilde{h}_{\alpha_1\alpha_2}(\boldsymbol{k_3}) \tilde{h}_{\alpha_3\alpha_4}(\boldsymbol{k_4})\\
    &=\frac{\ii e}{2\hbar} \sum_{\boldsymbol{k}} \sum_{\alpha_1...\alpha_4} \partial_y \tilde{h}_{\alpha_1\alpha_2}(\boldsymbol{k})\tilde{h}_{\alpha_3\alpha_4}(\boldsymbol{k}) [\{\partial_{k_1^x} -\partial_{k_2^x}\} \mathcal{A}_{\alpha_1\alpha_2\alpha_3\alpha_4}(\boldsymbol{k_1}, \boldsymbol{k_2})]|_{\boldsymbol{k_1=k_2=k}}
\end{aligned}
\label{eq:DM1}
\end{equation}
The first-order bare vertex contribution to $C^{\mathcal{P}}_{c_{\boldsymbol{k_1}, \alpha_1}^\dagger c_{\boldsymbol{k_1-p}, \alpha_2}; c_{\boldsymbol{k_2}, \alpha_3}^\dagger c_{\boldsymbol{k_2+p},\alpha_4}}(\omega = 0)$ is the summation of two diagrams, \cref{fig_sm1}~(e-f)
\begin{equation}
\footnotesize
\begin{aligned}
    \mathcal{B}_{\alpha_1\alpha_2\alpha_3\alpha_4}(\boldsymbol{k_1}, \boldsymbol{k_2},\boldsymbol{p}) &= \sum_{m_1...m_4} U_{\alpha_1m_1}^*(\boldsymbol{k_1}) U_{\alpha_2m_2}(\boldsymbol{k_1-p}) U_{\alpha_3m_3}^*(\boldsymbol{k_2})U_{\alpha_4m_4}(\boldsymbol{k_2+p}) \beta^{-2}\sum_{\omega_n, \omega_n'} \frac{1}{\ii \omega_n - \epsilon_{m_1}(\boldsymbol{k_1})} \frac{1}{\ii \omega_n - \epsilon_{m_2}(\boldsymbol{k_1-p})} \\
    & \frac{1}{\ii \omega_n' - \epsilon_{m_3}(\boldsymbol{k_2})}\frac{1}{\ii \omega_n' - \epsilon_{m_4}(\boldsymbol{k_2+p})} \frac{2}{N}(-V_{m_2m_4m_1m_3}(\boldsymbol{k_1-p}, \boldsymbol{k_2+p}, \boldsymbol{k_1-k_2-p}) +  V_{m_4m_2m_1m_3}(\boldsymbol{k_2+p, k_1-p,p}))\\
    &= -\frac{2}{N} \sum_{m_1,,,m_4} U_{\alpha_1m_1}^*(\boldsymbol{k_1}) U_{\alpha_2m_2}(\boldsymbol{k_1-p}) U_{\alpha_3m_3}^*(\boldsymbol{k_2})U_{\alpha_4m_4}(\boldsymbol{k_2+p}) \mathcal{P}[\frac{f(\epsilon_{m_1}(\boldsymbol{k_1})) - f(\epsilon_{m_2}(\boldsymbol{k_1-p}))}{\epsilon_{m_1}(\boldsymbol{k_1}) - \epsilon_{m_2}(\boldsymbol{k_1-p})}]\\
    &\quad\mathcal{P}[\frac{f(\epsilon_{m_3}(\boldsymbol{k_2})) - f(\epsilon_{m_4}(\boldsymbol{k_2+p}))}{\epsilon_{m_3}(\boldsymbol{k_2}) - \epsilon_{m_4}(\boldsymbol{k_2+p})}] (V_{m_2m_4m_1m_3}(\boldsymbol{k_1-p}, \boldsymbol{k_2+p}, \boldsymbol{k_1-k_2-p}) -  V_{m_4m_2m_1m_3}(\boldsymbol{k_2+p, k_1-p,p}))
\end{aligned}
\label{eq:bare_vertex_firstorder}
\end{equation}
Substituting \cref{eq:bare_vertex_firstorder} into \cref{eq:I_express} and then \cref{eq:OM} of the main text and using integration-by-parts, we have
\begin{equation}
\small
\begin{aligned}
    \delta \tilde{M}^{(2)}_{z} &= \frac{\ii e}{2\hbar N^4} \sum_{\boldsymbol{R_1}...\boldsymbol{R_4}} \sum_{\alpha_1...\alpha_4} \sum_{\boldsymbol{k_1k_2k_3k_4p}} (y_{\boldsymbol{R}_1, \alpha_1} - y_{\boldsymbol{R}_2, \alpha_2})
    (x_{\boldsymbol{R_1}, \alpha_1} + x_{\boldsymbol{R_2}, \alpha_2} - x_{\boldsymbol{R_3}, \alpha_3} - x_{\boldsymbol{R_4}, \alpha_4}) \\&
    \qquad\qquad  e^{-\ii \boldsymbol{k_1\cdot R}_{12, \alpha_1\alpha_2} - \ii \boldsymbol{k_2\cdot R}_{34, \alpha_3\alpha_4} - \ii \boldsymbol{p\cdot R}_{24, \alpha_2\alpha_4} + \ii \boldsymbol{k_3 \cdot R}_{12, \alpha_1\alpha_2} + \ii \boldsymbol{k_4\cdot R}_{34, \alpha_3\alpha_4}} \mathcal{B}_{\alpha_1\alpha_2\alpha_3\alpha_4}(\boldsymbol{k_1}, \boldsymbol{k_2}, \boldsymbol{p}) \tilde{h}_{\alpha_1\alpha_2}(\boldsymbol{k_3}) \tilde{h}_{\alpha_3\alpha_4}(\boldsymbol{k_4})\\
    &=\frac{\ii e}{2\hbar N^4} \sum_{\boldsymbol{R_1...R_4}} \sum_{\alpha_1...\alpha_4} \sum_{\boldsymbol{k_1k_2k_3k_4p}} \partial_{k_3^y} (\partial_{k_1^x} - \partial_{k_2^x} + 2\partial_{p^x}) [e^{-\ii \boldsymbol{k_1\cdot R}_{12, \alpha_1\alpha_2} - \ii \boldsymbol{k_2\cdot R}_{34, \alpha_3\alpha_4} - \ii \boldsymbol{p\cdot R}_{24, \alpha_2\alpha_4} + \ii \boldsymbol{k_3 \cdot R}_{12, \alpha_1\alpha_2} + \ii \boldsymbol{k_4\cdot R}_{34, \alpha_3\alpha_4}}]\\
    &\qquad\qquad\qquad \qquad\qquad\qquad\qquad\qquad\qquad\qquad\qquad\qquad\qquad\mathcal{B}_{\alpha_1\alpha_2\alpha_3\alpha_4}(\boldsymbol{k_1}, \boldsymbol{k_2}, \boldsymbol{p}) \tilde{h}_{\alpha_1\alpha_2}(\boldsymbol{k_3}) \tilde{h}_{\alpha_3\alpha_4}(\boldsymbol{k_4})\\
    &=\frac{\ii e}{2\hbar} \sum_{\boldsymbol{k_1k_2}}\sum_{\alpha_1...\alpha_4} \partial_y \tilde{h}_{\alpha_1\alpha_2}(\boldsymbol{k_1})\tilde{h}_{\alpha_3\alpha_4}(\boldsymbol{k_2}) [(\partial_{k_1^x}-\partial_{k_2^x}+2\partial_{p^x})  \mathcal{B}_{\alpha_1\alpha_2\alpha_3\alpha_4}(\boldsymbol{k_1},\boldsymbol{k_2},\boldsymbol{p})]|_{\boldsymbol{p}=0}
\end{aligned}
\label{eq:DM2}
\end{equation}
We finally consider the zeroth order expansion of $C^{\mathcal{P}}_{c_{\boldsymbol{k_1},\alpha_1}^\dagger c_{\boldsymbol{k_2}, \alpha_2}; c_{\boldsymbol{k_3},\alpha_3}^\dagger c_{\boldsymbol{k_4}, \alpha_4}^\dagger c_{\boldsymbol{k_5}, \alpha_4} c_{\boldsymbol{k_6}, \alpha_3}}(\omega = 0)$, which consists of four different connected contractions, as illustrated in \cref{fig_sm1}~(g).
Using Wick's theorem, the first line of \cref{fig_sm1}~(g) equals to
\begin{equation}
\begin{aligned}
    &- \delta_{\boldsymbol{k_1k_6}} \delta_{\boldsymbol{k_2k_3}}\delta_{\boldsymbol{k_4k_5}} \int_{0}^\beta d\tau G^{(0)}_{\alpha_3\alpha_1}(\boldsymbol{k_1}, -\tau) G^{(0)}_{\alpha_2\alpha_3}(\boldsymbol{k_2}, \tau) G^{(0)}_{\alpha_4\alpha_4}(\boldsymbol{k_4}, \eta)\\
    =& -\delta_{\boldsymbol{k_1k_6}} \delta_{\boldsymbol{k_2k_3}}\delta_{\boldsymbol{k_4k_5}} \int_{0}^\beta d\tau \beta^{-3}\sum_{\omega_n, \omega_n', \omega_n''} G^{(0)}_{\alpha_3\alpha_1}(\boldsymbol{k_1}, \ii \omega_n) G^{(0)}_{\alpha_2\alpha_3}(\boldsymbol{k_2}, \ii \omega_n') G^{(0)}_{\alpha_4\alpha_4}(\boldsymbol{k_4}, \ii \omega_n'') e^{\ii (\omega_n'-\omega_n)\tau} e^{\ii \omega_n'' \eta}\\
    =& -\delta_{\boldsymbol{k_1k_6}} \delta_{\boldsymbol{k_2k_3}}\delta_{\boldsymbol{k_4k_5}}  \beta^{-2}\sum_{\omega_n, \omega_n'}G^{(0)}_{\alpha_3\alpha_1}(\boldsymbol{k_1}, \ii \omega_n) G^{(0)}_{\alpha_2\alpha_3}(\boldsymbol{k_2}, \ii \omega_n) G^{(0)}_{\alpha_4\alpha_4}(\boldsymbol{k_4}, \ii \omega_n') e^{\ii \omega_n' \eta}
\end{aligned}
\end{equation}
The other three lines can be calculated similarly, the sum of the four lines is
\begin{equation}
\footnotesize
\begin{aligned}
    & \beta^{-2}\sum_{\omega_n, \omega_n'}\{-\delta_{\boldsymbol{k_1k_6}} \delta_{\boldsymbol{k_2k_3}}\delta_{\boldsymbol{k_4k_5}}  G^{(0)}_{\alpha_3\alpha_1}(\boldsymbol{k_1}, \ii \omega_n) G^{(0)}_{\alpha_2\alpha_3}(\boldsymbol{k_2}, \ii \omega_n) G^{(0)}_{\alpha_4\alpha_4}(\boldsymbol{k_4}, \ii \omega_n')
    +\delta_{\boldsymbol{k_1k_6}} \delta_{\boldsymbol{k_2k_4}}\delta_{\boldsymbol{k_3k_5}} G^{(0)}_{\alpha_3\alpha_1}(\boldsymbol{k_1}, \ii \omega_n) G^{(0)}_{\alpha_2\alpha_4}(\boldsymbol{k_2}, \ii \omega_n) G^{(0)}_{\alpha_4\alpha_3}(\boldsymbol{k_3}, \ii \omega_n')\\
    &+\delta_{\boldsymbol{k_1k_5}} \delta_{\boldsymbol{k_2k_3}}\delta_{\boldsymbol{k_4k_6}} G^{(0)}_{\alpha_4\alpha_1}(\boldsymbol{k_1}, \ii \omega_n)G^{(0)}_{\alpha_2\alpha_3}(\boldsymbol{k_2}, \ii \omega_n) G^{(0)}_{\alpha_3\alpha_4}(\boldsymbol{k_4}, \ii \omega_n') - \delta_{\boldsymbol{k_1k_5}}\delta_{\boldsymbol{k_2k_4}}\delta_{\boldsymbol{k_3k_6}} G^{(0)}_{\alpha_4\alpha_1}(\boldsymbol{k_1}, \ii \omega_n)G^{(0)}_{\alpha_2\alpha_4}(\boldsymbol{k_2}, \ii \omega_n) G^{(0)}_{\alpha_3\alpha_3}(\boldsymbol{k_3}, \ii \omega_n')\} e^{\ii \omega_n'\eta}
\end{aligned}
\label{eq:J_k}
\end{equation}
Substituting $G^{(0)}_{\alpha\beta}(\boldsymbol{k}, \ii \omega_n)= -\sum_{m} U_{\alpha m}(\boldsymbol{k})U_{\beta m}^*(\boldsymbol{k})\frac{1}{\ii \omega_n -\epsilon_m(\boldsymbol{k})}$ into \cref{eq:J_k} and then \cref{eq:J_express}, we have
\begin{equation}
\begin{aligned}
    &\mathcal{J}_{\alpha_1\alpha_2\alpha_3\alpha_4}(\boldsymbol{R_1}, \boldsymbol{R_2}, \boldsymbol{R_3}, \boldsymbol{R_4}) = \frac 1 {N^3} \sum_{\boldsymbol{k_1k_2k_3}} \sum_{m_1m_2m_3} \mathcal{P}[\frac{f(\epsilon_{m_1}(\boldsymbol{k_1})) - f(\epsilon_{m_2}(\boldsymbol{k_2}))}{\epsilon_{m_1}(\boldsymbol{k_1}) -\epsilon_{m_2}(\boldsymbol{k_2})}] f(\epsilon_{m_3}(\boldsymbol{k_3})) U_{\alpha_1 m_1}^*(\boldsymbol{k_1}) U_{\alpha_2m_2}(\boldsymbol{k_2})\\
    \{ &-U_{\alpha_3 m_2}^*(\boldsymbol{k_2}) U_{\alpha_3m_3}(\boldsymbol{k_3})U_{\alpha_4 m_3}^*(\boldsymbol{k_3}) U_{\alpha_4m_1}(\boldsymbol{k_1}) e^{-\ii \boldsymbol{k_1\cdot R}_{14, \alpha_1\alpha_4} +\ii \boldsymbol{k_2\cdot R}_{23, \alpha_2\alpha_3} + \ii \boldsymbol{k_3\cdot R}_{34, \alpha_3\alpha_4}}  \\
    &+U_{\alpha_3 m_2}^*(\boldsymbol{k_2}) U_{\alpha_3m_1}(\boldsymbol{k_1})U_{\alpha_4 m_3}^*(\boldsymbol{k_3}) U_{\alpha_4m_3}(\boldsymbol{k_3}) e^{-\ii \boldsymbol{k_1\cdot R}_{13, \alpha_1\alpha_3} +\ii \boldsymbol{k_2\cdot R}_{23, \alpha_2\alpha_3}} \\&+ U_{\alpha_3 m_3}^*(\boldsymbol{k_3}) U_{\alpha_3m_3}(\boldsymbol{k_3})U_{\alpha_4 m_2}^*(\boldsymbol{k_2}) U_{\alpha_4m_1}(\boldsymbol{k_1}) e^{-\ii \boldsymbol{k_1 \cdot R}_{14, \alpha_1\alpha_4} +\ii \boldsymbol{k_2 \cdot R}_{24, \alpha_2\alpha_4}}\\
    & - U_{\alpha_3 m_3}^*(\boldsymbol{k_3}) U_{\alpha_3m_1}(\boldsymbol{k_1})U_{\alpha_4 m_2}^*(\boldsymbol{k_2}) U_{\alpha_4m_3}(\boldsymbol{k_3}) e^{-\ii \boldsymbol{k_1\cdot R}_{13, \alpha_1\alpha_3} +\ii \boldsymbol{k_2\cdot R}_{24, \alpha_2\alpha_4} - \ii \boldsymbol{k_3 \cdot R}_{34, \alpha_3\alpha_4}}\}
\end{aligned}
\label{eq:J_real}
\end{equation}
Substituting \cref{eq:J_real} into \cref{eq:OM} of the main text and using integration-by-parts, we have
\begin{equation}
\small
\begin{aligned}
    \delta \tilde{M}_z^{(3)} =& +\frac{\ii e}{\hbar N}\sum_{\boldsymbol{k k'}} \sum_{m_1m_2m_3}\sum_{\alpha_1...\alpha_4} \partial_y \tilde{h}_{\alpha_1\alpha_2}(\boldsymbol{k}) \tilde{V}_{\alpha_3\alpha_4}(0) f(\epsilon_{m_3}(\boldsymbol{k'})) U_{\alpha_3m_3}(\boldsymbol{k'}) U_{\alpha_3m_3}^*(\boldsymbol{k'})\\
    &\qquad\qquad(\partial_{k_1^x} - \partial_{k_2^x}) \{\mathcal{P}[\frac{f(\epsilon_{m_1}(\boldsymbol{k_1})) - f(\epsilon_{m_2}(\boldsymbol{k_2}))}{\epsilon_{m_1}(\boldsymbol{k_1}) - \epsilon_{m_2}(\boldsymbol{k_2})}] U_{\alpha_4 m_1}(\boldsymbol{k_1}) U_{\alpha_1m_1}^*(\boldsymbol{k_1}) U_{\alpha_2 m_2}(\boldsymbol{k_2}) U_{\alpha_4 m_2}^*(\boldsymbol{k_2})\}|_{\boldsymbol{k_1 = k_2 = k}}\\
    & -\frac{\ii e}{\hbar N}\sum_{\boldsymbol{k k'}} \sum_{m_1m_2m_3}\sum_{\alpha_1...\alpha_4} \partial_y \tilde{h}_{\alpha_1\alpha_2}(\boldsymbol{k}) \tilde{V}_{\alpha_3\alpha_4}(\boldsymbol{k-k'}) f(\epsilon_{m_3}(\boldsymbol{k'})) U_{\alpha_3m_3}(\boldsymbol{k'}) U_{\alpha_4m_3}^*(\boldsymbol{k'})\\
    &\qquad\qquad(\partial_{k_1^x} - \partial_{k_2^x}) \{\mathcal{P}[\frac{f(\epsilon_{m_1}(\boldsymbol{k_1}))- f(\epsilon_{m_2}(\boldsymbol{k_2}))}{\epsilon_{m_1}(\boldsymbol{k_1}) - \epsilon_{m_2}(\boldsymbol{k_2})}] U_{\alpha_4 m_1}(\boldsymbol{k_1}) U_{\alpha_1m_1}^*(\boldsymbol{k_1}) U_{\alpha_2 m_2}(\boldsymbol{k_2}) U_{\alpha_3 m_2}^*(\boldsymbol{k_2})\}|_{\boldsymbol{k_1 = k_2 = k}}\\
    & +\frac{\ii e}{\hbar N}\sum_{\boldsymbol{k k'}} \sum_{m_1m_2m_3}\sum_{\alpha_1...\alpha_4} \partial_y \tilde{h}_{\alpha_1\alpha_2}(\boldsymbol{k}) \partial_x \tilde{V}_{\alpha_3\alpha_4}(0) f(\epsilon_{m_3}(\boldsymbol{k'})) U_{\alpha_3m_3}(\boldsymbol{k'}) U_{\alpha_3m_3}^*(\boldsymbol{k'})\\
    &\qquad\qquad\qquad\qquad\qquad\qquad\qquad\qquad\mathcal{P}[\frac{f(\epsilon_{m_1}(\boldsymbol{k}))- f(\epsilon_{m_2}(\boldsymbol{k}))}{\epsilon_{m_1}(\boldsymbol{k}) - \epsilon_{m_2}(\boldsymbol{k})}] U_{\alpha_4 m_1}(\boldsymbol{k}) U_{\alpha_1m_1}^*(\boldsymbol{k}) U_{\alpha_2 m_2}(\boldsymbol{k}) U_{\alpha_4 m_2}^*(\boldsymbol{k})\\
\end{aligned}
\label{eq:DM3}
\end{equation}
In summary, we have $\delta \tilde{M}_z = \delta \tilde{M}_z^{(1)} + \delta \tilde{M}_z^{(2)} + \delta \tilde{M}_z^{(3)}$, and since $\delta \tilde{M}_z$ is linear in $U$ we have $k_U = \frac{\partial \delta \tilde{M}_z}{\partial U}|_{U=0, \mu} = \delta \tilde{M_z}|_{U=1, \mu}$.
Note that the chemical potential $\mu$ is already absorbed into the definition of $\epsilon_m(\boldsymbol{k})$.

In practical calculations of $\delta \tilde{M}_z$ in a system with linear size $L$, the partial derivatives are replaced by finite differences. Define the scalar $q = 2\pi/L$ and vectors $\boldsymbol{q_x} = q\boldsymbol{\hat{x}}, \boldsymbol{q_y} = q\boldsymbol{\hat{y}}$, we replace $[\{\partial_{k_1^x} -\partial_{k_2^x}\} \mathcal{A}_{\alpha_1\alpha_2\alpha_3\alpha_4}(\boldsymbol{k_1}, \boldsymbol{k_2})]|_{\boldsymbol{k_1=k_2=k}}$ with $(2q)^{-1}\{ \mathcal{A}_{\alpha_1\alpha_2\alpha_3\alpha_4}(\boldsymbol{k_1+q_x}, \boldsymbol{k_2-q_x}) - \mathcal{A}_{\alpha_1\alpha_2\alpha_3\alpha_4}(\boldsymbol{k_1-q_x}, \boldsymbol{k_2+q_x})\}$, and replace $\partial_y \tilde{h}_{\alpha\beta}(\boldsymbol{k})$ with $(2q)^{-1} \{ \tilde{h}_{\alpha\beta}(\boldsymbol{k+q_y}) - \tilde{h}_{\alpha\beta}(\boldsymbol{k-q_y})\}$ in \cref{eq:DM1}. Similar rules for partial derivatives are also applied to \cref{eq:DM2,eq:DM3}.

\end{document}